\documentclass[twocolumn,showpacs,preprintnumbers,amsmath,amssymb,aps,prl]{revtex4}
\usepackage{graphicx}
\begin{document}

\title{Spontaneous Transverse Response and Amplified Switching in Superconductors with Honeycomb Pinning Arrays} 
\author{C. Reichhardt and   
 C.J. Olson Reichhardt} 
\affiliation{
Theoretical Division,
Los Alamos National Laboratory, Los Alamos, New Mexico 87545}

\date{\today}
\begin{abstract}
Using numerical simulations, 
we show that a novel spontaneous transverse response
can appear when a longitudinal drive is applied 
to type-II superconductors with honeycomb pinning arrays 
in a magnetic field near certain filling fractions.
This response is generated by dynamical symmetry breaking that
occurs at fields away from commensurability. 
We find a coherent strongly amplified transverse switching effect
when an additional transverse ac drive is applied.  
The transverse ac drive can also be used to 
control switching in the longitudinal velocity response.
We discuss how these effects could be used to create new types of
devices such as current effect transistors. 
\end{abstract}
\maketitle

\vskip2pc
There have been  
extensive studies 
on superconducting systems with patterned pinning arrays which show
commensurability effects when the vortex density matches 
the pinning site density \cite{Baert,Karapetrov,Reichhardt,Peeters}.  
Various properties of the vortex transport can be controlled by
adjusting the shape and geometry of the pinning arrays, 
producing
vortex channeling  \cite{Velez},
dynamical transitions \cite{Olson,Rinke}, 
fluxon ratchets \cite{Janko}, 
reversible ratchets \cite{Vicent,Morelle}, 
and vortex cellular automata \cite{Hastings}.
These studies indicate that vortices interacting with patterned 
substrates may lead to a new field of fluxtronics or 
microelectronic devices based on the controlled motion of vortices.
Further, vortices interacting with periodic pinning arrays 
also exhibit a number of 
collective dynamic behaviors which are
important to the broader field of nonequilibrium physics. 

Few studies 
have been performed 
on vortex dynamics in honeycomb pinning arrays
since it was assumed that the vortex behavior   
in this pinning geometry would be similar to 
that found in square and
triangular pinning arrays. 
In this Letter we demonstrate that in fact,
honeycomb pinning arrays produce new types 
of phenomena that do not occur in triangular or square pinning arrays.   
In particular,
an effective dimerization or higher order $n$-merization of the  
interstitial vortices 
in the honeycomb pinning arrays for certain field ranges  
results in a novel 
{\it  spontaneous transverse response} (STR)
to a longitudinal drive. 
Normally, an applied current 
produces a perpendicular Lorentz force on superconducting
vortices, which move and generate a voltage 
drop parallel
to the current.
We show that the honeycomb system can also produce a voltage 
drop {\it perpendicular} to the applied current.     
For magnetic fields away from commensuration, 
the vortex flow {\it dynamically organizes} to a STR state. 
At the matching fields, the 
STR
arises due to symmetry  breaking in the ground state.    
The vortex $n$-mer states are 
very similar to the colloidal molecular crystal
states studied for colloids interacting with periodic 
optical trap arrays \cite{Bechinger}. 
We specifically consider vortices in superconductors; however,
we expect 
similar phases to occur whenever there is a spontaneous 
symmetry breaking of effective $n$-mer states for 
particles on periodic substrates. 

We simulate a two-dimensional system containing $N_v$ vortices and
$N_p$ pinning sites with periodic boundary conditions in the
$x$ and $y$ directions.  The vortex 
density is 
$B=N_v\phi_0/L^2$,  
where $L=24\lambda$ is the system size in units of the
penetration depth $\lambda$ and
$\phi_0=h/2e$ is the elementary flux quantum. 
The motion of a single vortex 
is given by 
the following 
overdamped equation:
\begin{equation} 
\eta \frac{d{\bf R}_{i}}{dt} = {\bf F}^{vv}_{i} + {\bf F}^{p}_{i} + 
{\bf F}^{dc} + {\bf F}^{ac}  + {\bf F}^{T}_{i} .
\end{equation}
Here 
${\bf R}_{i}$ is the location of vortex $i$, 
$\eta = \phi_{0}^2d/2\pi\xi^2\rho_{N}$ is the damping constant, 
$d$ is the thickness of the superconducting sample, 
$\xi$ is the coherence length, and $\rho_{N}$  is the normal state resistivity.
The pairwise vortex-vortex interaction
force is ${\bf F}^{vv}_{i} = \sum^{N_{v}}_{i\neq j}f_{0}K_{1}(R_{ij}/\lambda){\hat {\bf R}}_{ij}$  
where $K_{1}$ is the modified Bessel function, 
$R_{ij}=|{\bf R}_i-{\bf R}_j|$, 
${\hat {\bf R}}_{ij}=({\bf R}_i-{\bf R}_j)/R_{ij}$, 
and $f_{0} = \phi_{0}^2/(2\pi\mu_{0}\lambda^3)$.
The pinning force ${\bf F}^{p}_{i}$ arises from parabolic traps 
of radius $r_p=0.3\lambda$ and strength $f_p=1.0f_0$ arranged in
a honeycomb lattice with
${\bf F}^{p}_{i}=-\sum_{k=1}^{N_p}f_pR_{ik}r_p^{-1}\Theta((R_p-R_{ik})/\lambda){\bf {\hat R}}_{ik}.$
Here $R_{ik}=|{\bf R}_i-{\bf R}_k|$ is
the distance between vortex $i$ and pin $k$,
${\hat {\bf R}}_{ik}=({\bf R}_i-{\bf R}_k)/R_{ik}$, and
$\Theta$ is the Heaviside step function. 
The dc longitudinal driving force ${\bf F}^{dc} = F^{dc}f_0{\hat {\bf R}}^{L}$, 
where we take the longitudinal direction ${\bf {\hat R}}^{L}$
to be ${\bf {\hat R}}={\bf {\hat x}}$.
This force mimics the Lorentz force created by an applied current.   
We also consider the effect of adding a transverse ac drive
${\bf F}^{ac} = F^{ac}f_0\sin(\omega t){\hat{\bf R}}^{Tr}$, 
where $\omega=10^{-7}f_0/\eta$ and 
${\bf {\hat R}}^{Tr}={\bf {\hat y}}$ is the transverse direction. 
The thermal force
${\bf F}_i^{T}$ arises from Langevin kicks and has the properties
$\langle F^{T}_i(t)\rangle = 0$ 
and 
$\langle F^{T}_i(t)F^{T}_j(t^{\prime})\rangle =2\eta k_{B}T\delta_{ij}\delta(t - t^{\prime})$.  
Vortex positions are initialized with simulated annealing.

\begin{figure}
\includegraphics[width=3.5in]{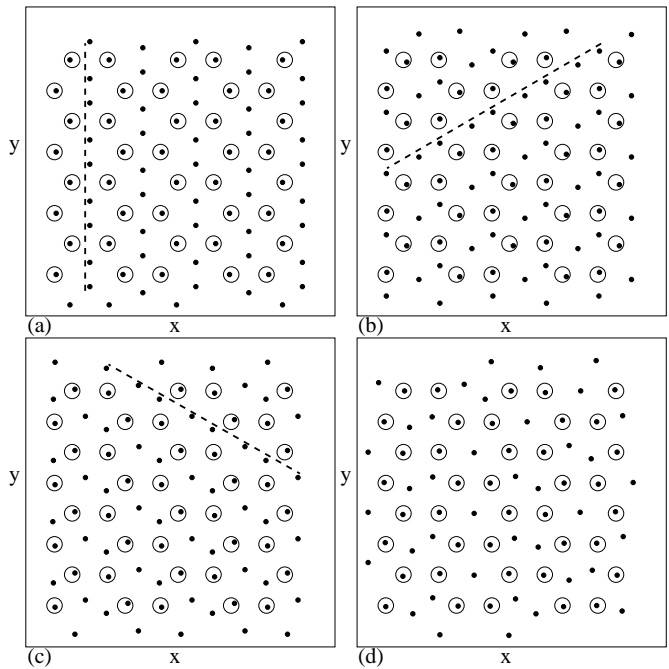}
\caption{
Vortex positions (black dots) and pinning site locations (open
circles) in a $20\lambda \times 20\lambda$ section of the sample
at $F^{dc} = 0$.
(a,b,c): Three different realizations
of $B/B_\phi=2.0$ with interstitial dimers aligned (a) in the
$y$-direction; (b) at $+30^\circ$ to the $x$-direction;
(c) at $-30^\circ$ to the $x$-direction. 
The dashed lines indicate the alignment direction.
(d) $B/B_{\phi} = 1.77$ with no overall dimer alignment. 
}
\end{figure}

In Fig.~1 we illustrate the vortex and pinning site positions
in a sample with $F^{dc}=0$, $F^{ac}=0$,
and $T = 0$. 
At $B/B_\phi=2.0$, shown in Fig.~1(a,b,c), 
where $B_\phi$ is the field at which $N_v=N_p$,
the large interstitial space of each plaquette in the
honeycomb pinning lattice traps
two interstitial vortices which form an effective rigid dimer with a
director field that can point in one of three degenerate directions, 
as indicated by the dashed lines.
Neighboring dimers interact through
an effective quadrupole interaction 
which causes
the dimers to align into one of three degenerate ground states
\cite{Olson2}. 
For fillings $1.5 < B/B_\phi < 2.0$ and $2.0 < B/B_\phi < 2.5$,
the dimer alignment is disrupted and there is {\it no} global symmetry 
breaking in the ground state. 
Figure 1(d) shows an example at
$B/B_{\phi} = 1.77$ where the 
dimer ordering is lost.

\begin{figure}
\includegraphics[width=3.5in]{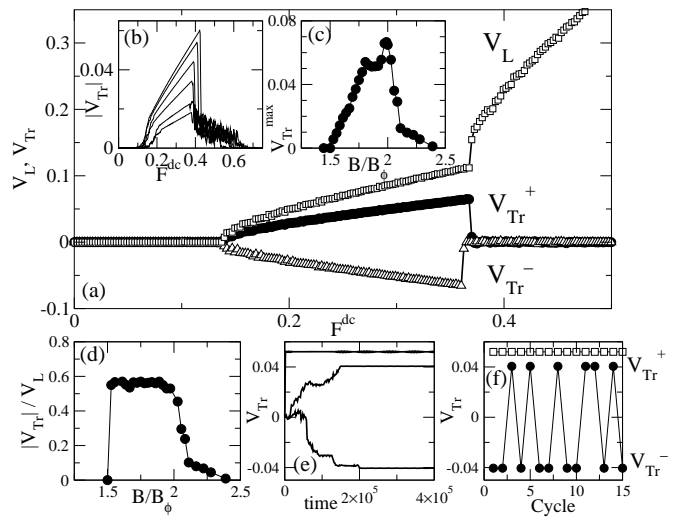}
\caption{
(a) Transverse velocities $V_{Tr}^{\pm}$ and longitudinal velocity $V_{L}$ 
versus 
$F^{dc}$ for the sample in Fig.~1 at 
$B/B_\phi=2.0$. 
(b) $|V_{Tr}|$ vs $F^{dc}$ for the same system at
$B/B_{\phi} = 1.58$, 1.61, 1.67, 1.72, 1.77, and $1.81$, from bottom to top.
(c) $V_{Tr}^{max}$
versus $B/B_{\phi}$. 
(d) $|V_{Tr}|/V_{L}$ versus $B/B_{\phi}$ 
at fixed $F^{dc} = 0.225$.      
(e) Time series of $V_{Tr}$ for $F^{dc}=0.3$ at $B/B_{\phi}=2.0$ (top curve)
and $B/B_{\phi}=1.81$ (two lower curves).
(f) $V_{Tr}$ during consecutive cycles of $F^{dc}$ from 0 to $F^{dc}=0.3$ at
$B/B_\phi=2.0$ (open squares) and $B/B_\phi=1.81$ (filled circles).
}
\end{figure}

We next apply a dc drive $F^{dc}$
in the longitudinal direction and measure 
the vortex velocity ${\bf v}$ to obtain the
longitudinal velocity  
$V_L=\sum_i^{N_v}{\bf v}_i \cdot {\bf {\hat R}}^L$
and the transverse velocity  
$V_{Tr}=\sum_i^{N_v}{\bf v}_i \cdot {\bf {\hat R}}^{Tr}$.
In Fig.~2(a) we plot $V_{L}$ and $V_{Tr}$ versus $F^{dc}$ for 
the system in Fig.~1 at $B/B_\phi=2.0$.
For $F^{dc} < 0.14$
the system is pinned, while
for $0.14 \le F^{dc} < 0.37$ there is 
a finite longitudinal velocity $V_L$ accompanied by 
a finite transverse 
response $V_{Tr}$
that can be in either the positive ($V_{Tr}^+$) or negative 
($V_{Tr}^-$) transverse direction. 
The trajectories of the moving vortices for the $V_{Tr}^+$ and
$V_{Tr}^-$ states
are shown in Figs.~3(a) and (b),
respectively. 
When $F^{dc}>F^{dc}_c$, where $F^{dc}_c$ is the critical drive at which
vortices in the pinning sites begin to depin, the symmetry breaking is
lost.
At $B/B_\phi=2.0$, $F^{dc}_c=0.37$, and above this drive,
$|V_{Tr}|$ drops abruptly to zero and
$V_{L}$ rapidly increases.  
The curve with $V_{Tr}^+$ 
was obtained 
by starting from the ground state in Fig.~1(a) at $T = 0$ 
while the curve 
with 
$V_{Tr}^-$ was generated starting from the same ground
state but applying a finite but small temperature $T=0.2T_m$, where $T_m$ is
the vortex lattice melting temperature.
If we repeat the finite temperature simulation with different random seeds,
we are equally likely to observe $V_{Tr}^+$ or $V_{Tr}^-$.
If the initial ground state already has a global symmetry preferred positive or negative
orientation with respect to the transverse direction, as in
Figs. 1(b) and (c), the transverse response is in the
same direction indicated by the dashed lines in Fig. 1(b,c).
The appearance of a transverse velocity at $B/B_{\phi} = 2.0$ is not a manifestation 
of a dynamical symmetry breaking since the 
dimer alignment symmetry
is already broken as in Fig.~1.
In contrast, for the 
fields $1.5 < B/B_{\phi} < 2.0$ and $2.0 <  B/B_{\phi} < 2.5$ there is no  
global symmetry breaking in the ground states; however, a 
{\it dynamical symmetry breaking}   
occurs when the vortices organize into a dynamical phase
resembling the states in Figs.~3(a,b)
after
a transient period of time 
during which 
the vortices move in
both the $V_{Tr}^{+}$ and $V_{Tr}^-$ directions. 

\begin{figure}
\includegraphics[width=3.5in]{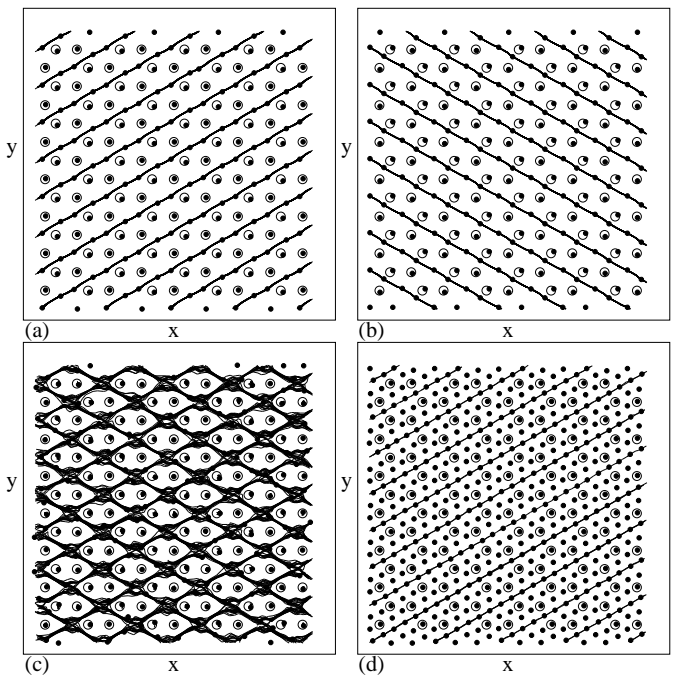}
\caption{
Vortex positions (black dots), pinning site locations (open circles), 
and trajectories (lines) over a constant time interval. 
(a) The $V_{Tr}^+$ dynamically broken symmetry state from Fig.~2(a) 
at $F^{dc}=0.225$.
(b) The $V_{Tr}^-$ state at $F^{dc} = 0.225$.
(c) A sample with $B/B_{\phi} = 2.11$ at $F^{dc}=0.4$
where complete symmetry breaking is lost.  
(d) The $V_{Tr}^+$ state in a sample with $B/B_\phi=4.5$ and $F^{dc} = 0.15$.
}
\end{figure}

In 
Fig.~2(b) 
we plot $|V_{Tr}|$ as a function of 
$F^{dc}$ at $B/B_{\phi} = 1.58$, 1.61, 1.67, 1.72, 1.77, and $1.81$ 
showing the appearance of the dynamical broken symmetry state.  
In 
Fig.~2(e) 
we  plot $V^{Tr}$ vs time for $F^{dc} = 0.3$ at 
$B/B_{\phi} = 2.0$ 
and $B/B_{\phi} = 1.81$, 
where the latter curves were
generated from samples in slightly different initial states.   
At $B/B_{\phi} = 2.0$, the vortices immediately move in the 
broken symmetry direction of 
the ground state, which in this case is $V_{Tr}^+$ shown in Fig.~1(b). 
For $B/B_{\phi} = 1.81$, the initial motion is symmetric and 
only develops into a $V_{Tr}^+$ or $V_{Tr}^-$
broken symmetry state over time.
The system may even fluctuate significantly in one direction before 
locking into the other direction.   
This shows that the moving states can have global symmetry breaking even
when the equilibrium ground state does not.
Due to the initial symmetric transient moving state,
the memory of the pinned state configuration is lost at incommensurate fields 
but retained at commensurate fields 
when $F^{dc}$ is repeatedly cycled from 
zero into the moving broken symmetry state and back again. 
In 
Fig.~2(f),
the plot of $V_{Tr}$ at $B/B_{\phi} = 2.0$ and 1.81 
during consecutive cycles of $F^{dc}$ 
shows that
the sign of the symmetry breaking switches
randomly between cycles
at the incommensurate field. 

In 
Fig.~2(c) 
we plot $V_{Tr}^{max}$, the maximum value of $|V_{Tr}|$ obtained
when varying $F^{dc}$, 
versus $B/B_{\phi}$. 
The overall maximum amount of transverse
motion occurs at $B/B_\phi=2.0$.  The fraction of
the 
velocity
that is in the longitudinal direction
$|V_{Tr}|/V_{L}$
is shown in
Fig.~2(d) 
for fixed $F^{dc}=0.225$
as a function of $B/B_{\phi}$. 
For complete symmetry breaking, the vortices move at $\pm 30^{\circ}$ from
the longitudinal axis so we expect
$|V_{Tr}|/V_L=\tan(30^\circ)=0.577$, as seen for $1.5 < B/B_\phi \le 2.0$ in
Fig.~2(d).
For $B/B_{\phi} > 2.0$, both 
$V_{Tr}^{max}$ and $|V_{Tr}|/V_L$ drop rapidly. 

The same type of symmetry breaking flow 
occurs at higher commensurate states such as
$B/B_\phi=4.5$ where elongated trimers can form, as illustrated in 
Fig.~3(d).
This is accompanied by spontaneous dynamical symmetry breaking 
over a range of fields.
We have also considered the role of temperature. 
Previous work \cite{Olson2} indicated that the
interstitial vortices melt at a well defined 
temperature $T_{m}$;  
using parameters appropriate for Nb crystals gives 
$T_m \approx 8.5$ K.
We find that our results are robust for all $T < T_{m}$. 
In Fig.~4(d) we plot $|V_{Tr}|$ vs $T/T_{m}$ at $B/B_{\phi} = 2.0$
and $F^{dc} = 0.3$, showing 
that $V_{Tr}$ drops sharply with increasing $T$ just below $T/T_{m} = 1.0$.

\begin{figure}
\includegraphics[width=3.5in]{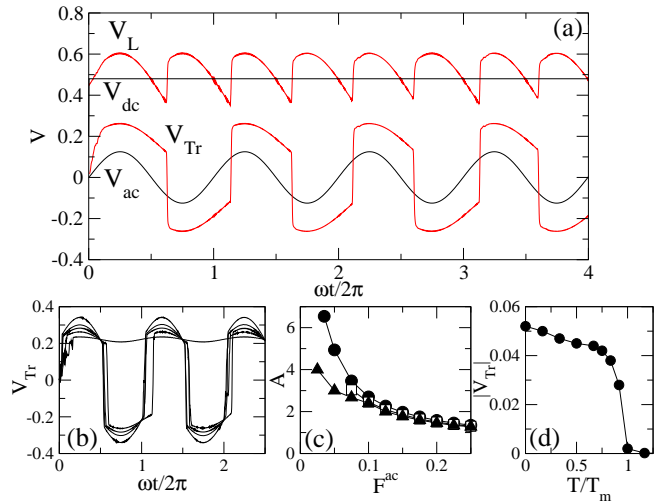}
\caption{
(a) $V_{Tr}(t)$ (thick lower line) and $V_L(t)$ (thick upper line), 
with the analytic values of $V_{ac}(t)$ (thin lower line)
and $V_{dc}(t)$ (thin upper line) shown for comparison.
$V_L$ and $V_{dc}$ have been shifted up by 0.2 for presentation purposes.
Time is given in units of $2\pi/\omega$.
Here $B/B_\phi=2.0$, $F^{ac}=0.125$, and $F^{dc}=0.28$.
(b) Time series of $V_{Tr}$ for a system with 
$B/B_\phi=1.77$, $F^{dc} = 0.28$, and varied ac drive amplitude of
$F^{ac}=0.15$, 0.1, 0.075, 0.05, 0.035, and 0.02, from 
top to bottom on the positive response side.  
(c) The amplification factor $A$ versus $F^{ac}$ 
for samples at $F^{dc} = 0.28$ with $B/B_{\phi} = 1.61$ (filled circles), 
$2.0$ (open squares), and $2.11$ (filled triangles).  
(d) $|V_{Tr}|$ vs $T/T_{m}$ at $B/B_{\phi} = 2.0$
and $F^{dc} = 0.3$.
}
\end{figure}
        
Since the dynamical symmetry-broken states are bistable, 
it is natural to ask whether it is possible to induce a 
switching behavior between the two states. 
To address this, we consider the effect of adding a 
transverse oscillating drive $F^{ac}$ to a system 
with $B/B_\phi=2.0$ moving under a fixed
longitudinal drive $F^{dc}=0.28$.
Experiments have already been conducted in superconductors
with periodic pinning arrays where two orthogonal driving currents 
were simultaneously applied and the 
transverse and longitudinal responses were simultaneously measured 
\cite{Gonzalez}. 
In Fig.~4(a) we plot the time series
$V_{Tr}(t)$ and $V_{L}(t)$ 
in a system with $F^{ac}=0.125$.
Here the ac drive induces a periodic switching
between the positive and negative 
transverse response states $V_{Tr}^+$ and $V_{Tr}^-$. 
We find that this abrupt switching effect    
persists for $1.5 < B/B_{\phi} \leq 2.0$.        
Similar switching appears over a wide range of ac driving
amplitudes, as illustrated in Fig.~4(b) where we plot
$V_{Tr}(t)$ at $B/B_{\phi} = 1.77$ and $F^{dc}=0.28$
for different values of $F^{ac}$. 
We observe a novel response in the longitudinal direction in 
which the value of $V_{L}(t)$ varies by up to 50\% during each ac drive period.
There is also a switching effect in $V_{L}(t)$ 
which accompanies each switch in $V_{Tr}(t)$.
This phenomenon is reminiscent of the 
current effect transistor 
found in transversely driven charge density waves 
where 
a transverse force can 
be used to control the longitudinal response \cite{LeoR}.   

There is an amplification of the transverse ac response $V_{Tr}(t)$
over the value expected based only on the magnitude of the ac input
driving signal.
To illustrate this, in Fig.~4(a) we plot
$V_{ac}(t) \equiv N_mF^{ac}(t)/\eta$,
the transverse ac velocity 
produced when
the 
$N_m=N_v-N_p$
interstitial vortices move only 
in response to the ac driving force,  
along with
$V_{dc} \equiv N_mF^{dc}/\eta$,
the longitudinal velocity 
of the interstitial vortices 
under only the dc drive.
Fig.~4(a) indicates that $|V_{Tr}(t)|>|V_{ac}(t)|$.
The relative amplification $V_{Tr}(t)/V_{ac}(t)$
can be increased by lowering $F^{ac}$; however, for a finite 
$F^{dc}$ there is a threshold value $F^{ac}_{c}$ below which the system
no longer switches between $V_{Tr}^+$ and $V_{Tr}^-$.
For example, in Fig.~4(b), $F^{ac}=0.02$ is below the threshold $F^{ac}_c$, and
thus the response stays locked in the $V_{Tr}^+$ direction.  
We quantify the amplification of the transverse ac response using
$A=V_{Tr}\eta/(N_mF^{ac})$, where $V_{Tr}$ is the amplitude of the
transverse response, such that for $A=1.0$, there is no amplification.
The plot of $A$ vs $F^{ac}$ in Fig.~4(c)
for $B/B_{\phi} = 1.61$, $2.0$, and $2.11$
shows that for a fixed dc drive, the amount of 
ac amplification that occurs depends on both $F^{ac}-F^{ac}_c$ 
and $F^{dc}_c-F^{dc}$.
The closer the ac and dc drives are to the respective critical thresholds,
the larger the amplification.
Since $F^{dc}_c$ is maximized at $B/B_{\phi} = 2.0$, for fixed $F^{dc}=0.28$
a larger amplification can be obtained at $B/B_\phi \ne 2.0$ than at
$B/B_\phi=2.0$, as shown in Fig.~4(c).
For the parameters used in Fig.~4, 
the frequencies for a typical superconductor are between 10 and 30 Hz.
If $\omega$ is increased, the sharp switching response can still be achieved
by increasing $F^{ac}$ until 
$\omega \sim 20$ to 60 KHz,
at which
point $F^{ac}_c$ crosses above the depinning threshold for the pinned
vortices and the dynamical symmetry breaking is completely lost.    

Many of the switching features, 
including bistability in the 
transverse response and the abrupt switching effect induced by the ac drive, 
are analogous to semiconductors
and may be useful for creating logic devices. 
The strong coupling between the longitudinal and transverse
responses indicates that a transverse ac drive could be used to       
control longitudinal switching, while
the amplification effect implies
that
very small ac inputs can induce large response changes in certain regimes.
Our results should apply 
to any system of repulsively interacting 
particles confined by honeycomb pinning arrays. We have observed
a similar dynamical symmetry breaking flow for colloids on 
triangular substrates 
\cite{Colloid}; however, the dynamical symmetry breaking for the
vortex system presented here 
is much more robust and occurs for a considerably wider range of parameters 
and particle fillings.      

In summary, we have demonstrated that 
vortices in superconductors with honeycomb pinning arrays can exhibit 
a novel transverse response when a longitudinal drive is 
applied for certain ranges of fields where the vortices in the large 
interstitial regions of the pinning array form effective dimer or 
higher order $n$-mer states. At commensurate fields, the 
symmetry is broken in the ground state, while at incommensurate
fields, there is a dynamical symmetry breaking.  
If a transverse ac drive is added to the system,
a pronounced transverse switching response 
occurs, as well as an amplification of the transverse ac signal. 
There is a threshold ac drive required to induce
the switching which becomes very small when the longitudinal drive is close
to the value above which the transverse response disappears.
The transverse ac drive can be used to  modulate the
longitudinal response of the system as well. 
We discuss how this effect may be useful for creating
new types of fluxon based devices.   

This work was carried out under the auspices of the 
NNSA of the 
U.S. DoE
at 
LANL
under Contract No.
DE-AC52-06NA25396.

\end{document}